\documentclass[12pt,preprint]{aastex}

\slugcomment{accepted for ApJL}

\shorttitle{Quasar \ion{Fe}{2}/\ion{Mg}{2} Ratio}
\shortauthors{Verner  and Peterson }

\begin{document}


\title{
Elucidating the Correlation of the Quasar \ion{Fe}{2}/\ion{Mg}{2} Ratio with
Redshift}


\author{E. M.~Verner\altaffilmark{1,2} and B. A.~Peterson\altaffilmark{3}}

\altaffiltext{1}{IACS/Dept. of Physics, Catholic University of America, Washington, DC, 20064}
\altaffiltext{2}{Laboratory for Astronomy and Solar Physics, Code 681, Goddard Space 
Flight Center Greenbelt MD 20771}
\altaffiltext{3}{Mt. Stromlo Observatory, Research School of Astronomy and
Astrophysics, The Australian National University,
Weston Creek Post Office, A.C.T. 2611, AUSTRALIA}

\email{kverner@fe2.gsfc.nasa.gov, ~peterson@mso.anu.edu.au}



\begin{abstract}

Interpretation of the \ion{Fe}{2}(UV)/\ion{Mg}{2} emission ratios from quasars 
has a major cosmological motivation. Both Fe and Mg are produced 
by short-lived massive stars. 
In addition, Fe is produced by accreting white dwarf supernovae somewhat after star 
formation begins. Therefore, we expect that the Fe/Mg ratio will
gradually decrease with redshift. 
We have used data from the Sloan Digital Sky Survey to 
explore the dependence of the 
\ion{Fe}{2}(UV)/\ion{Mg}{2}  ratio on redshift and on 
luminosity in the redshift range of
$0.75< z< 2.20$, and
we have used predictions from our 830-level model 
for the \ion{Fe}{2} atom in photoionization calculations 
to interpret our findings.

We have split the quasars into several groups based upon the value of 
their \ion{Fe}{2}(UV)/\ion{Mg}{2}  emission ratios,  
and then checked to see how the fraction of quasars in 
each group varies with the increase of redshift. 
We next examined the luminosity dependence of the  
\ion{Fe}{2}(UV)/\ion{Mg}{2} ratio,  and we found that beyond a threshold of
\ion{Fe}{2}(UV)/\ion{Mg}{2}~=~5, and $M_{2500} < -25\rm\ mag$, the 
\ion{Fe}{2}(UV)/\ion{Mg}{2} ratio increases
with luminosity, as predicted by our model.

We interpret our observed variation of the 
\ion{Fe}{2}(UV)/\ion{Mg}{2} ratio with 
redshift as
a result of the correlation of redshift 
with luminosity in a magnitude limited quasar sample.

\end{abstract}


%% subject header.  Objects should be in the appropriate "individual"
%% headers (e.g. quasars: individual, stars: individual, etc.) with the
%% additional provision that the total number of headers, including each
%% individual object, not exceed six.  The \objectname{} macro, and its
%% alias \object{}, is used to mark each object.  The macro takes the object
%% name as its primary argument.  This name will appear in the paper
%% and serve as the link's anchor in the electronic edition if the name
%% is recognized by the data centers.  The macro also takes an optional
%% argument in parentheses in cases where the data center identification
%% differs from what is to be printed in the paper.

\keywords{atomic processes---line: formation---methods: numerical---quasars: emission lines}


\section{Introduction}

Recently, the number of observational efforts focused on  \ion{Fe}{2}(UV)/\ion{Mg}{2} 
emission ratio measurements has increased due to the potential use of these ratios to trace 
star formation history.   Both Fe and Mg are produced 
by short lived massive stars (SN type II), and additional
Fe is produced by accreting white dwarf 
supernovae (SN type Ia) sometime later.
Thus, one expects that the Fe/Mg ratio
is low at high redshift, and gradually increases with decreasing redshift,
with the increase starting at a redshift of about 4, corresponding to an age 
for the Universe of 1.5 Gyr (Hamann and Ferland 1993; Yoshii et al. 1998),
or even as early as a redshift of 6 if star formation in the extreme environment
of a quasar central region begins at $z \ge 10$ (Matteucci \& Recchi 2001).

Although many independent ways to measure \ion{Fe}{2}(UV)/\ion{Mg}{2} ratios 
observed in the UV to IR range have been developed and applied to different data sets, 
the common conclusion is that the
\ion{Fe}{2}(UV)/\ion{Mg}{2} ratio shows a large scatter at all redshifts, and little
evolution with redshift.
(Kwan \& Krolik 1981; Wills et al. 1985; Kinney et al. 1991; Kawara et al. 1996;
Thompson et al. 1999; Iwamuro et al.  2002; Dietrich et al. 2002;
Freudling et al. 2003; Barth et al. 2003;
Dietrich et al. 2003; and Maiolino et al. 2003). 
Furthermore, it is usually assumed that there is a
linear dependence between the \ion{Fe}{2}(UV)/\ion{Mg}{2} and Fe/Mg ratios
because the ionization
potentials of Mg (7.65 eV) and Fe (7.87 eV) are nearly the same.

In our previous study, we showed that the \ion{Fe}{2}(UV)/\ion{Mg}{2}  
ratios are more sensitive to other physical properties of the emitting 
region  than to abundance (Verner et al. 2003; 2004). If the 
physical conditions are different in the Broad Line Regions (hereafter, BLR) of different quasars, 
the resulting scatter of \ion{Fe}{2}(UV)/\ion{Mg}{2}  
ratios obscures any dependence on abundance. 
Thus, it is important to explain the origin of the observed scatter before attempting to
derive the Fe/Mg relative abundance in quasars.

Although the prominent emission lines observed in quasars  generally 
allow us to evaluate the metalicity of galactic nuclei, to link specifically the
\ion{Fe}{2} emission with Fe abundance is not a simple task. 
Compared to many other ions, the \ion{Fe}{2}  ion has a very rich spectrum due to its 
half-filled 3d-shell. As a result the \ion{Fe}{2}(UV) band from 
2200$-$3000\AA\  [hereafter, \ion{Fe}{2}(UV))] can contain hundreds of strong lines. 
The \ion{Fe}{2}(UV)/\ion{Mg}{2} emission ratio is therefore heavily 
affected by many \ion{Fe}{2} lines but by only  the two \ion{Mg}{2} doublet lines  
at ~2800\AA. Nevertheless, in order to measure the Fe/Mg abundance ratio, 
obtaining the \ion{Fe}{2}(UV)/\ion{Mg}{2} ratio 
is unavoidable. 

As quasar spectra (line intensities and continuum) are heavily affected by 
\ion{Fe}{2} emission, it is possible to learn more about BLRs in quasars 
by studying how the \ion{Fe}{2} originates. To achieve such a goal we have 
constructed an 830-level model for the
\ion{Fe}{2} atom and investigated how \ion{Fe}{2}(UV)/\ion{Mg}{2} ratios 
vary with changes in hydrogen density, microturbulence and abundance. 
The model also predicts that the \ion{Fe}{2}(UV)/\ion{Mg}{2} ratio strongly 
depends upon the ionizing flux of the central source.

In this Letter we have investigated if there is any dependence of 
\ion{Fe}{2}(UV)/\ion{Mg}{2} ratios with redshift. For the first time our
approach combines model predictions 
with measurements [provided by Iwamuro (2004)]
of the \ion{Fe}{2}(UV)/\ion{Mg}{2} ratios of quasars
in the extended redshift range, $0.75 < z < 2.20$, for quasars in the Sloan Digital Sky Survey 
(hereafter, SDSS)\footnote{http ://www.sdss.org/}.

\section{\ion{Fe}{2}(UV)/\ion{Mg}{2} Dependence on Redshift}
The \ion{Fe}{2}\  spectrum has a rich and complicated mixture of forbidden and 
permitted lines. The pioneering work on the term analysis of singly ionized iron 
(Russell 1926) reported 61 energy levels. Since work on \ion{Fe}{2} energy 
measurements is still not completed (Johansson 1978, 2004)
it is not  surprising that the very first attempts to explain \ion{Fe}{2} in 
BLRs were not able to reproduce large \ion{Fe}{2}(UV)/\ion{Mg}{2} 
ratios when assuming solar abundance (Kwan \& Krolik 1983; Wills, Netzer, \& Wills 1985).

Our 830-level model for \ion{Fe}{2} atom in photoionization calculations is able 
to account for non-abundance factors (e.g. microturbulence, hydrogen density, and ionizing flux). 
This model is a natural extension (up to 14.1 eV) of our earlier model for Fe$^{+}$,
which included 371 energy levels below 11.6 eV (Verner et al. 1999). 
The increase in the number of transitions from 68,638 to 344,035 is largely due to the increased 
density of energy levels at higher energy. The \ion{Fe}{2}(UV)/\ion{Mg}{2} ratios 
have been investigated assuming ranges of iron abundances (1, 5, 10 in solar units) 
and microturbulence $v_{turb}~= 0, 1, 10, 100$~km~s$^{-1}$.
Based on our model predictions we conclude that the \ion{Fe}{2}(UV)/\ion{Mg}{2} ratio is 
less sensitive to abundance than the \ion{Fe}{2}(UV)/\ion{Fe}{2}(Opt) ratio
(where \ion{Fe}{2}(Opt) is at 4000$-$6000\AA) (Verner et al. 2003, 2004).

After we found that the abundance is not the strongest factor that determines the
\ion{Fe}{2}(UV)/\ion{Mg}{2} ratio in BLRs of quasars, calculations 
were repeated for a wide range of physical conditions, assuming solar abundance,  and
constant $v_{turb}~=~5$~km~s$^{-1}$, hydrogen density range
{$n_H$ }= 10 $^{9.5} - 10^{13.0}$cm$^{-3}$, total column
density, $N_{H}$ = 10$^{24 }~$cm$^{-2}$, and the flux of
hydrogen ionizing photons at the illuminated face  10$^{17.5} - 10^{22.0}~$photons cm$^{-2}$
s$^{-1}$. We employ the characteristic AGN
continuum described in Korista et al. (1997), which consists of a UV bump
peaking near 44 eV, a $f_{\nu} \propto \nu^{-1}$ X-ray power law, and a UV to
optical spectral index, $a_{ox}=  -1.4$. Figure 1 presents the results of model 
calculations of the \ion{Fe}{2}(UV)/\ion{Mg}{2} ratio as a function of density 
and ionizing photon flux. Although solar abundance is assumed throughout,  
our model predicts that the \ion{Fe}{2}(UV)/\ion{Mg}{2} may reach values 
as large as  40 under conditions of high density and high ionizing photon flux.
Most observed values lie in the range $2<r<5$, with only a small
fraction falling in the range $10<r<20$.
The \ion{Fe}{2}(UV)/\ion{Mg}{2} ratio should be relatively constant 
at low density and weak  ionizing photon flux, but should rise steeply with  
ionizing photon flux
beyond $\Phi$ = 10$^{20.0}~$photons cm$^{-2}$s$^{-1}$ and {$n_H$ }$~>~$ 10 $^{11} $cm$^{-3}$.

Our large \ion{Fe}{2} model used in photoionization calculations
predicts that the \ion{Fe}{2}(UV)/\ion{Mg}{2} ratios can have the same value over 
a wide range of physical conditions (Fig. 1). Therefore the typical observed
\ion{Fe}{2}(UV)/\ion{Mg}{2} ratio of $ \sim 4$
may arise from
different excitation and density regimes.

The SDSS database (Schneider et al. 2003) includes 11,677
objects within $0.75 < z < 2.2$.
Iwamuro (2004) was able to measure the
\ion{Fe}{2}(UV)/\ion{Mg}{2} 
ratio for 10,670 of these quasars, and these
are included in the sample under study. 
The smallest and the largest selected redshifts are $z_{min} = 0.7485$ 
and $z_{max} = 2.1964$, correspondingly. 
The median redshift of the sample  
is $z_{med}~=~1.4573$.

In Fig. 1, we see that at high flux levels, (log $\Phi > 20.5$), the  \ion{Fe}{2}(UV)/\ion{Mg}{2} ratio
behaves differently in the high (log $n_H >11$) and low (log $n_H <11$) density regimes. 
At high density, collisions enhance the \ion{Fe}{2}(UV) emission,
while at low density, the \ion{Fe}{2}(UV) emission drops relative to the \ion{Mg}{2} emission.
Very large values of the  \ion{Fe}{2}(UV)/\ion{Mg}{2} ratio, those greater than 10,
are expected only at high luminosity
and high density.
At lower flux levels (log $\Phi < 20.5$),
the \ion{Fe}{2}(UV)/\ion{Mg}{2} ratio shows little density dependence, and
we can generally conclude that  the \ion{Fe}{2}(UV)/\ion{Mg}{2} ratio depends  on the ionizing flux.

We divide quasars into several groups depending on the value of ionizing flux, 
keeping the \ion{Fe}{2}(UV)/\ion{Mg}{2} ratios within a relatively small range: 
1) \ion{Fe}{2}(UV)/\ion{Mg}{2}$ < 2$ group shows almost no dependence on hydrogen density,
and is mainly due to low ionizing flux;
2) $2 < $ \ion{Fe}{2}(UV)/\ion{Mg}{2} $ < 5$ group includes most quasars; 
3) $5 < $ \ion{Fe}{2}(UV)/\ion{Mg}{2} $ < 7.5$ group has a ratio a little bit higher 
than the typical value;
4) $7.5 < $ \ion{Fe}{2}(UV)/\ion{Mg}{2} $ < 10$ ratios in this group in previous 
studies were usually interpreted as having increased Fe abundance;
5) \ion{Fe}{2}(UV)/\ion{Mg}{2} $ > 10$ group with large ratios that are relatively rare.
 
The measured \ion{Fe}{2}/\ion{Mg}{2} ratio depends strongly on the fitting technique. 
High values of 
the  \ion{Fe}{2}/\ion{Mg}{2}  ratio result when the  \ion{Fe}{2} contribution to the \ion{Mg}{2}  doublet
is taken into account using 
the quasar template from Wills, Netzer, Wills (1985). 
Results based on the I~ZW~1 template 
(Vestergaard \& Wilkes 2001, Dietrich et al. 2002) 
underestimate contribution of \ion{Fe}{2} to the 
\ion{Mg}{2}  doublet (see Verner
et al. 2003; 2004 for more detailed discussion).

Iwamuro et al. (2002), have developed a parameterized fitting procedure 
to obtain the \ion{Fe}{2}(UV)/\ion{Mg}{2} ratio. The \ion{Fe}{2}(UV)/\ion{Mg}{2} ratio 
is defined  over 2150-3300~\AA ~to include the majority 
of \ion{Fe}{2}(UV) emission and to avoid the feature below 2150~\AA. This feature is  
possibly due to the contribution of  a Comptonized accretion disk (Zheng et al. 1998). 
Iwamuro et al. applied their method to independently obtained data in the wide redshift range, 
$0< z< 5.3$, [Kinney et al. 1991 and 
HST/FOS archival data ($z< 0.17$) , SDSS archival data ($0.75< z < 2.29$), 
%% changed higest to high
Thomson et al. 1999 data ($3.1 < z< 4.7$) and the high redshift sample ($4.4 < z< 5.3$)].
As a result of such examination they concluded that  \ion{Fe}{2}(UV)/\ion{Mg}{2}  
emission ratios in quasars show large scatter from
1 to 20 with little evidence of evolution of the \ion{Fe}{2}(UV)/\ion{Mg}{2} ratio with redshift,
and that luminosity does not have a large effect on the  \ion{Fe}{2}(UV)/\ion{Mg}{2} ratio,
except for extremely bright objects.

We have re-examined the redshift dependence of the \ion{Fe}{2}(UV)/\ion{Mg}{2} ratios 
measured by Iwamuro (2004)
in the SDSS quasars over  the range $0.75<z<2.20$. We compute the fraction of quasars 
with various \ion{Fe}{2}(UV)/\ion{Mg}{2} ratios in each redshift bin ($\Delta z~\sim 0.2$). 
We find that the fractions are constant at low redshift, but show an  increase in the relative numbers 
of quasars with \ion{Fe}{2}(UV)/\ion{Mg}{2}$ > 5$ 
at redshifts greater than $z~\sim 1.8$ (Fig. 2). This goes in the opposite direction of that
predicted by the supernova enrichment scenario, which predicts that the
excess Mg production in the early Universe would cause the Fe/Mg to
decrease with redshift.

In any survey which is magnitude limited, the most luminous objects will be
seen over the entire survey volume, while the least luminous objects will
only be seen nearby. This induces a correlation between luminosity and
redshift. 
We adopt a cosmology with {H$_0=70\rm\ km\ s^{-1}\ Mpc^{-1}$
and $\Omega_m =0.3$, $\Omega_\lambda = 0.7$ for the estimation of the 
absolute AB magnitude at rest 2500~\AA, $M_{2500}.$
The smallest and the
largest  $M_{2500}$ in the sample are 
$M^{min}_{2500} = -21.67\rm\ mag$ and $M^{max}_{2500}= -28.64\rm\ mag$.  
The median luminosity of the sample  
is $M^{med}_{2500} = -24.86\rm\ mag$.

For each \ion{Fe}{2}(UV)/\ion{Mg}{2} ratio, we plot the mean absolute luminosity of
the quasars in each redshift bin vs. redshift, and find a very strong
correlation between redshift and luminosity (Fig. 3).

We next examine the luminosity dependence of the 
\ion{Fe}{2}(UV)/\ion{Mg}{2} ratio in the SDSS quasars over  
the range $0.75<z<2.20$. We compute the fraction of quasars with
various \ion{Fe}{2}(UV)/\ion{Mg}{2} ratios in each luminosity bin. 
We find the fractions are constant at low luminosity, but show an increase 
in the relative numbers of quasars with 
\ion{Fe}{2}(UV)/\ion{Mg}{2}$~> 5$ at high luminosity, $M_{2500}<-25\rm\ mag$. 
We have found that beyond a threshold,
the  \ion{Fe}{2}(UV)/\ion{Mg}{2} ratio increases with quasar luminosity (Fig. 4).
This behavior is entirely consistent with our model (Fig. 1).
At low flux levels, $\log \Phi < 19$, increasing luminosity
has little effect on the  \ion{Fe}{2}(UV) /\ion{Mg}{2} ratio, while
at higher flux levels, $\log \Phi > 19$, the ratio steeply increases.

The much larger number of atomic levels for \ion{Fe}{2} atom used in the 
photoionization calculations provides an enhanced accuracy to account for the 
radiation field and enables us to explain the large \ion{Fe}{2}(UV) /\ion{Mg}{2} 
ratios better than in any previous efforts.
The complicated level structure of \ion{Fe}{2} ion 
leads to a more rapid increase of \ion{Fe}{2} emission with increasing of ionizing flux.  
More \ion{Fe}{2} levels become populated and more lines contribute to the total 
\ion{Fe}{2} emission. 
As no iron overabundance is  needed to explain the large 
\ion{Fe}{2}(UV) /\ion{Mg}{2} ratios that are observed,
we have used solar 
abundances throughout our study (Verner et al. 2003, 2004). 

Observations of \ion{Fe}{2}(Opt) are needed to constrain whether  the discovered change 
of \ion{Fe}{2}(UV) /\ion{Mg}{2} ratio with luminosity also reflects a change in other physical conditions
(e.g. hydrogen density) in emitting regions of BLRs.

\section{Conclusions}

While the extensive quantitative  comparisons between 
\ion{Fe}{2}(UV)/\ion{Mg}{2} ratio measurements and model 
predictions  are needed, it is clear that our model predictions are
in a general agreement with our observational results. Although it has been suggested
to use the \ion{Fe}{2}(UV)/\ion{Mg}{2} ratio to trace evolution, the physical
processes forming lines in the  \ion{Fe}{2}(UV) band  means that the \ion{Fe}{2}(UV)/\ion{Mg}{2}
ratio tells us
more about the central ionizing source than about abundances. Thus, we can   
only convert an \ion{Fe}{2}(UV)/\ion{Mg}{2} ratio into an Fe/Mg ratio when  a measurement 
of the \ion{Fe}{2}(Opt) 
band has also been obtained.

Our analysis demonstrates that  the observed \ion{Fe}{2}  spectra 
can be produced by a wide range of  ionizing flux regimes.
The statistical approach applied to physically 
distinguished  \ion{Fe}{2}(UV)/\ion{Mg}{2} ratio groups will 
provide a more efficient way to improve our understanding of  BLR in quasars
than the universal template approach. 

Due to complexity and richness of the \ion{Fe}{2} spectra in quasars, 
only the comparison between model predictions and  extensive observations 
make it possible to provide accurate measurements of changes of physical 
conditions and abundances in quasars with redshift. It is possible that the change we 
see is due to not only luminosity but to the hydrogen density changes in
the BLR with redshift as well. 
Wide wavelength coverage in the observations is needed for a successful understanding.

We find an increase in the \ion{Fe}{2}(UV)/\ion{Mg}{2} ratio with redshift,
while evolutionary models predict a decrease.
The \ion{Fe}{2}(UV)/\ion{Mg}{2} ratio is not sensitive to abundance changes
(see Fig. 4, Verner et al. 2003), but it is strongly affected by luminosity.
We have found that the relative numbers of quasars with a high 
\ion{Fe}{2}(UV)/\ion{Mg}{2} ratio increases with luminosity, 
as predicted by our model.

We conclude
that the apparent change
in the \ion{Fe}{2}(UV)/\ion{Mg}{2} ratio with redshift is simply a result of 
the correlation of
redshift with luminosity in the magnitude limited quasar sample, and is not
produced by a change in abundance.

%% The displaymath environment will produce the same sort of equation as
%% the equation environment, except that the equation will not be numbered
%% by LaTeX.


%% Included in this acknowledgments section are examples of the
%% AASTeX hypertext markup commands. Use \url without the optional [HREF]
%% argument when you want to print the url directly in the text. Otherwise,
%% use either \url or \anchor, with the HREF as the first argument and the
%% text to be printed in the second.

\acknowledgments

We are very grateful to Dr. Fumihide Iwamuro for providing us with
his measurements of the \ion{Fe}{2}/\ion{Mg}{2} emission ratios
in the spectra of SDSS quasars.

We are happy to acknowledge discussions with Prof. Kawara and his group at
the University of Tokyo regarding the chemical evolution of Fe/Mg, the effect
of luminosity on \ion{Fe}{2} emission, and the observation of these phenomena.
Our analysis provides an independent observational and theoretical confirmation
that, for  the brightest quasars,
the  luminosity of the central source affects the \ion{Fe}{2}/\ion{Mg}{2} emission
ratio (Iwamuro et. al 2002).

This research of EV has been supported, through NSF grant 
(NSF - 0206150) to CUA  and in part by visiting  program at 
Mt. Stromlo Observatory, ANU.

Funding for the Sloan Digital Sky Survey (SDSS) has been provided by the 
Alfred P. Sloan Foundation, the Participating Institutions, the National 
Aeronautics and Space Administration, the National Science Foundation, 
the U.S. Department of Energy, the Japanese Monbukagakusho, and the Max 
Planck Society. The SDSS Web site is http://www.sdss.org/.
The SDSS is managed by the Astrophysical Research Consortium (ARC) for 
the Participating Institutions. The Participating Institutions are The 
University of Chicago, Fermilab, the Institute for Advanced Study, the 
Japan Participation Group, The Johns Hopkins University, Los Alamos 
National Laboratory, the Max-Planck-Institute for Astronomy (MPIA), 
the Max-Planck-Institute for Astrophysics (MPA), New Mexico State 
University, University of Pittsburgh, Princeton University, the United 
States Naval Observatory, and the University of Washington.

\clearpage

%% the file, uncomment the \plotone, \plottwo, and
%% If you need a layout that cannot be achieved with \plotone or
%% \plottwo, you can invoke the graphicx package directly with the
%% \includegraphics command or use \plotfiddle. For more information,
%% please see the tutorial on "Using Electronic Art with AASTeX" in the
%% documentation section at the AASTeX Web site,
%% http://www.journals.uchicago.edu/AAS/AASTeX.
%%
%% The examples below also include sample markup for submission of
%% supplemental electronic materials. As always, be sure to check
%% the instructions to authors for the journal you are submitting to
%% for specific submissions guidelines as they vary from

%% This example uses \plotone to include an EPS file scaled to
%% 80% of its natural size with \epsscale. Its caption
%% has been written to indicate that additional figure parts will be
%% available in the electronic journal.

\figcaption [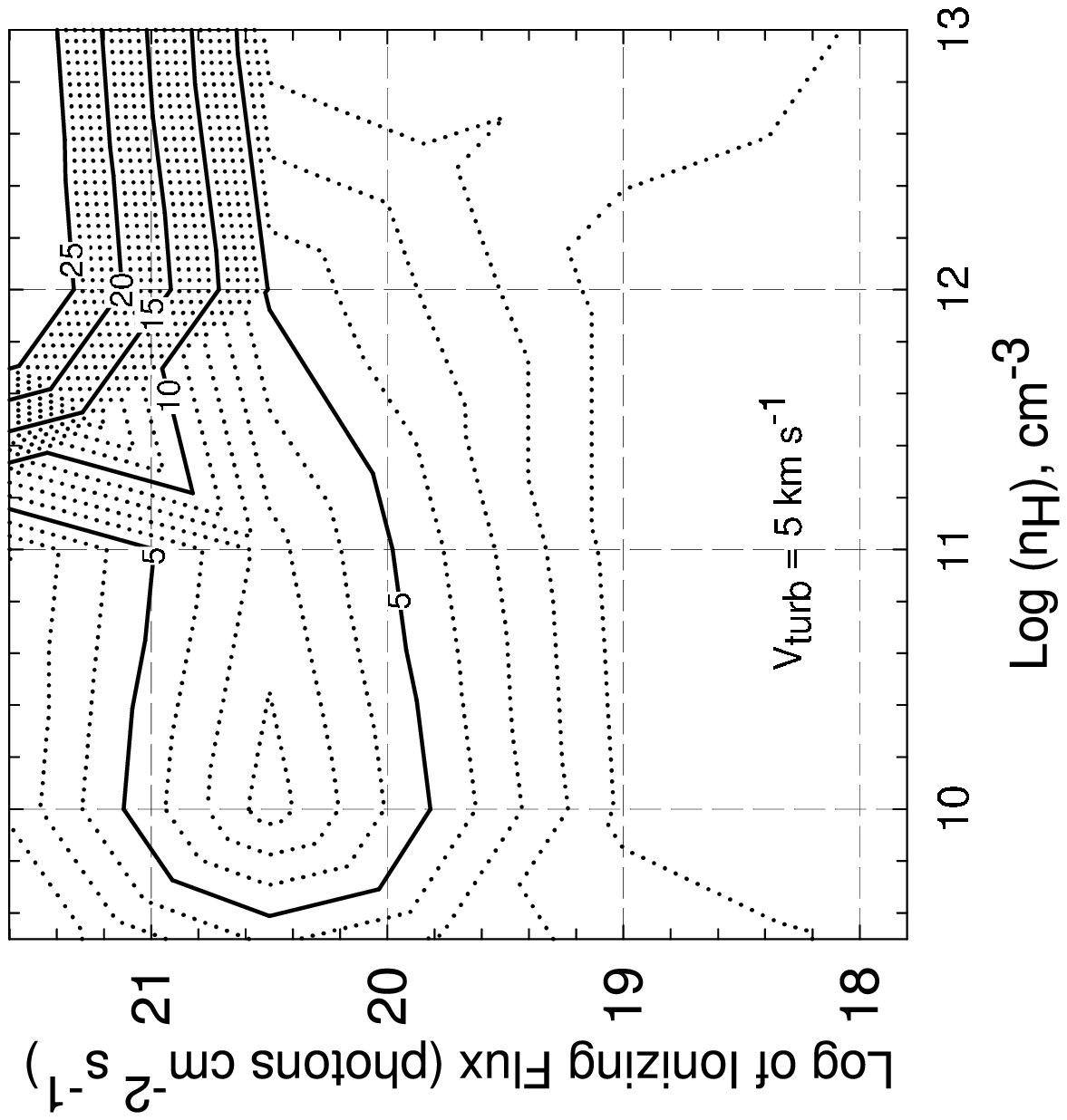]{The \ion{Fe}{2}(UV)/\ion{Mg}{2} ratios predicted using our
830-level model for Fe$^+$ in a quasar BLR as a function of the hydrogen 
density $n_H$ and flux of hydrogen-ionizing photons. The chemical abundances are solar, 
the cloud column density is $10^{24}$~cm$^{-3}$, and $v_{turb}~=~5$~km~s$^{-1}$.}

\figcaption [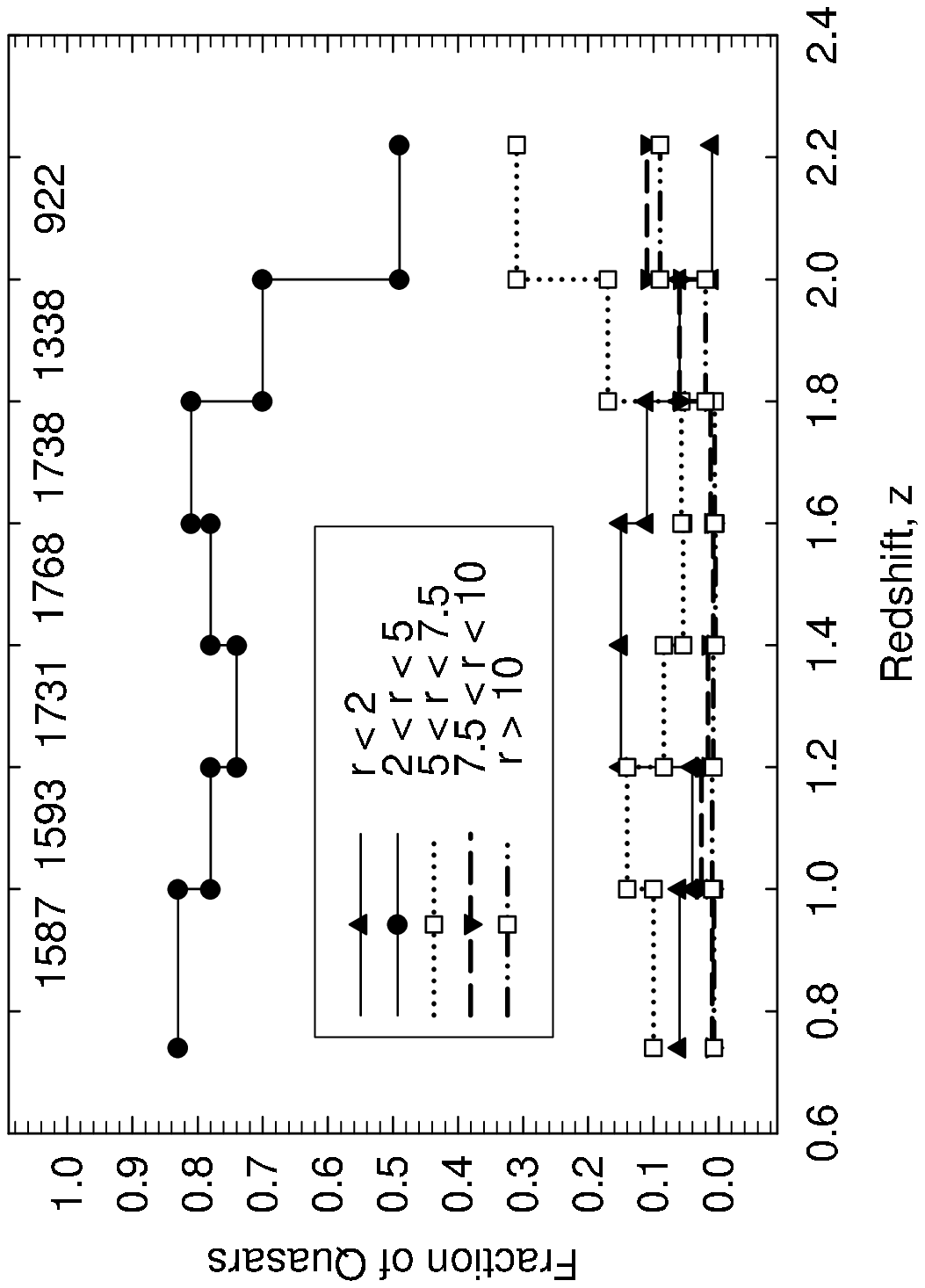]{The fraction of quasars in each redshift bin for various
\ion{Fe}{2}(UV)/\ion{Mg}{2} ratio groups. The numbers on the top show the total
number of quasars in each bin. The quasar sample comprises those quasars 
from from the SDSS in the redshift range 
$0.75<z<2.20$.}

\figcaption [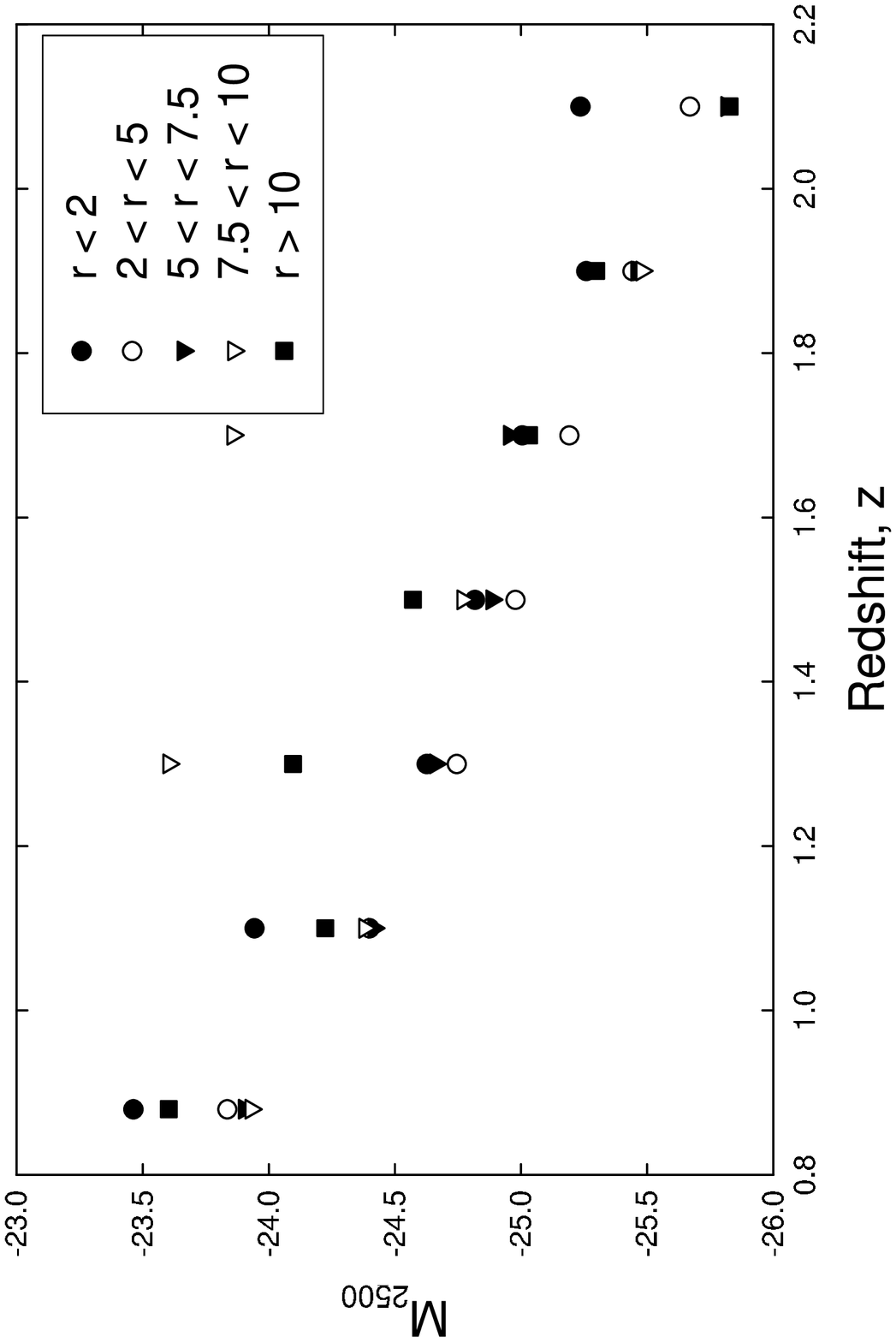]{The mean absolute luminosity for each \ion{Fe}{2}(UV)/\ion{Mg}{2} group 
of the quasars in each redshift bin vs. redshift.The quasar sample comprises those quasars 
from from the SDSS in the redshift range 
$0.75<z<2.20$.}

\figcaption [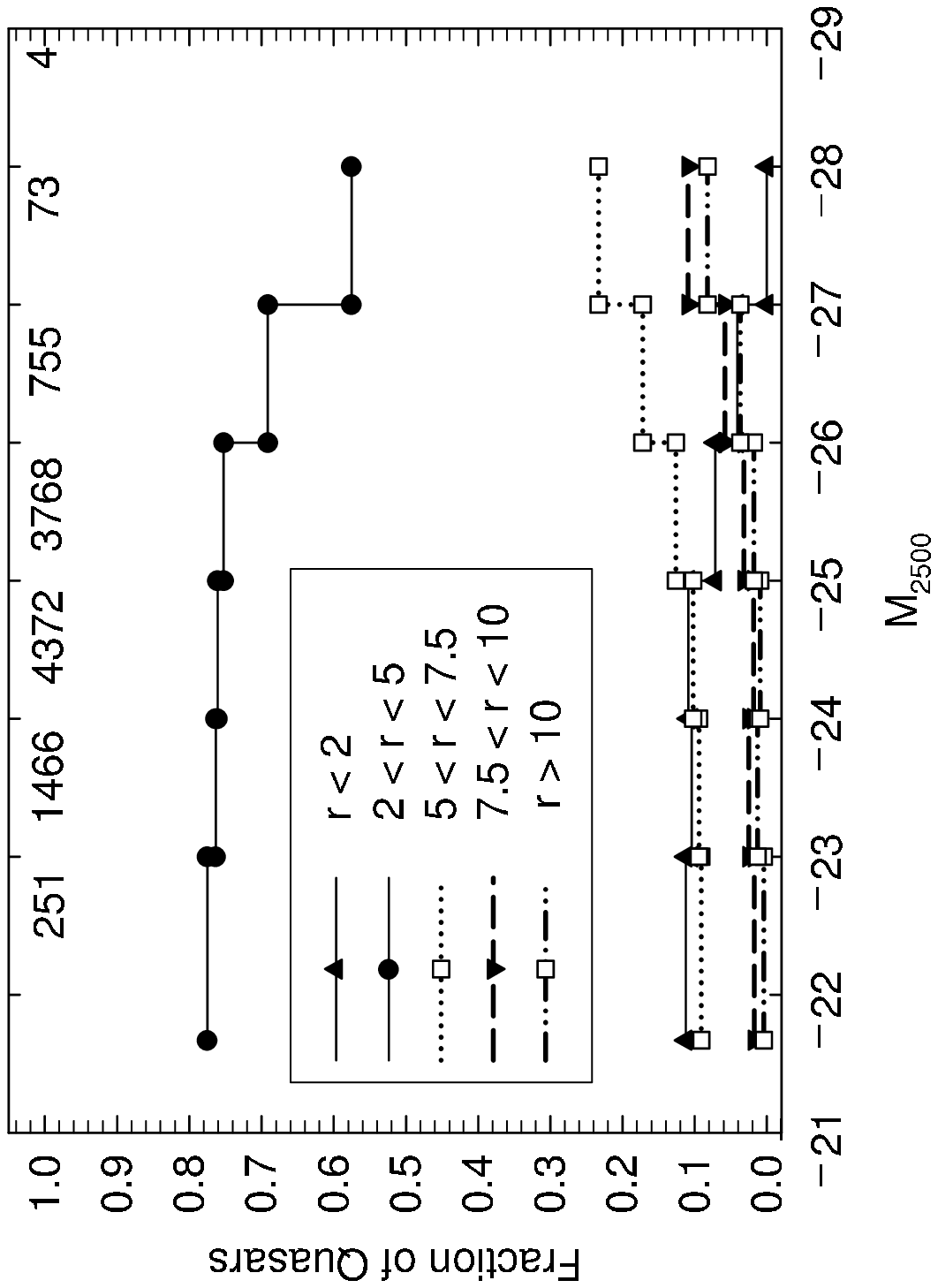]{The fraction of quasars in each luminosity bin for various
\ion{Fe}{2}(UV)/\ion{Mg}{2} ratio groups. The numbers on the top show the total 
number of quasars in each bin. The last bin with only 4 quasars has been ignored.
The quasar sample comprises those quasars from from the SDSS in the redshift range 
$0.75<z<2.20$.}

%%%UCP%%%
\newpage
%\rotatebox{-90}{
\epsscale{1.0}
\plotone{f1.eps}

%\rotatebox{-90}{
\epsscale{1}
\plotone{f2.eps}

%\rotatebox{-90}{
\epsscale{0.8}
\plotone{f3.eps}

%\rotatebox{-90}{
\epsscale{1.0}
\plotone{f4.eps}


\begin{thebibliography}{}
\bibitem[Barth et al. (2003)]{Bar03} Barth, A., Martini, P., Nelson, C. H.,
\& Ho, L. C. 2003, \apj, 594, 95
\bibitem[Dietrich et al. (2002)]{Die02} Dietrich, M., Appenzeller, I.,
Vestergaard, M. ., \& Wagner, S. J. 2002, \apj, 564, 581
\bibitem[Dietrich et al. (2003)]{Die03} Dietrich, M., Hamann, F., \&
Vestergaard, M. 2003, \apj, 596, 817
\bibitem[Freudling et al. (2003)]{Freu03} Freudling, W.,  Corbin, M., \&
Korista, K. 2003, \apj, 587, L67
\bibitem[Hamann \& Ferland (1993)]{Ham93} Hamann, F., \& Ferland, G. 1993,
\apj, 418, 11
\bibitem[Iwamuro et al. (2002)]{Iwa02} Iwamuro, F., Motohara, K., Maihara,
T., Kimura, M., Yoshii, Y., \& Doi, M. 2002, \apj,
565, 63
\bibitem[Iwamuro (2004)]{Iwa04} Iwamuro, F. 2004, private communication
\bibitem[Johansson (1978)]{Joh78} Johansson, S. 1978, Physica Scripta, 18, 217
\bibitem[Johansson (2004)]{Joh04} Johansson, S. 2004, preliminary database,
Lund Observatory
\bibitem[Kawara et al. (1996)]{Kaw96} Kawara, K., Murayama, T., 
Taniguchi, Y., \& Arimoto, N. 1996, \apj, 470, L85
\bibitem[Kinney et al. (1991)]{Kin91} Kinney, A. L., Bohlin, R. C., Blades, J. C., 
\& York, D. G. 1991, \apjs, 75, 645
\bibitem[Korista et al. (1997)]{Kor97} Korista, K., Baldwin, J., Ferland,
G., \& Verner, D. 1997, \apjs, 108,   401
\bibitem[Kwan \& Krolik (1981)]{Kwa81} Kwan, J. \& Krolik, J. 1981, \apj, 250, 478
\bibitem[Maiolino et al. (2003)]{Mai03} Maiolino, R., Juarez, Y., Mujica,
R., Nagar, N. M., \& Oliva, E. 2003, \apj, 596,   155
\bibitem[Matteucci \& Recchi (2001)]{MR01} Matteucci, F. \& Recchi, S. 2001, \apj, 558 351
\bibitem[Schneider et al.(2003)]{Sch03} Schneider, D., et al. 2003 \apj, 126, 2579
\bibitem[Thompson, Hill, \& Elston (1999)]{THE99} Thompson, K. L., Hill, 
G. J., \& Elston, R. 1999, \apj, 515, 487 
\bibitem[Verner et al. (1999)]{Ver99} Verner E. M., Verner, D. A., Korista,
K.T., Ferguson, J. W., Hamann, F., \& Ferland, G. J.
1999, \apjs, 120, 101
\bibitem[Verner (2000)]{Vern00} Verner, E. 2000, Ph.D. thesis, Univ. of
Toronto
\bibitem[Verner et al. (2003)]{Ver03} Verner, E., Bruhweiler, F., Verner,
D., Johansson, S., , \& Gull, T. 2003, \apj, 592, L59
\bibitem[Verner et al. (2004)]{Ver04} Verner, E., Bruhweiler, F., Verner,
D., Johansson, S., Kallman, T., \& Gull, T. 2004, \apj, submitted
\bibitem[Vestergaard \& Wilkes 2001]{VW01} Vestergaard, M., \& Wilkes, B. J. 2001, ApJSS, 134, 1
\bibitem[Wheeler et al. (1989)]{Whe89} Wheeler, J. C., Sneden, C., \&
Truran, J. W. 1989, \araa, 27, 279
\bibitem[Wills et al. (1985)]{WNW85} Wills, B. J., Netzer, H., \& Wills, D.
1985, \apj, 288, 94
% \bibitem[Yoshii et al. (1996)]{Yos96} Yoshii, Y., Tsujimoto, T., \& Nomoto, K. 1996, \apj, 462, 266
\bibitem[Yoshii et al. (1998)]{Yos98} Yoshii, Y., Tsujimoto, T., \& Kawara, K. 1998, \apj, 507, L113
\bibitem[Zheng et al. (1998)]{Zhe98} Zheng, W., Kriss, G. A., Telfer, R. C., 
Grimes, J. P., \& Davidsen, A. F. 1998,
\apj, 492, 855
\end{thebibliography}
\end{document}